\documentclass[preprint2]{proto}
\usepackage{times}
\newcommand{\refs}{\par\noindent\hangindent=1pc\hangafter=1}
\begin{document}

\title{\textbf{\LARGE The Structure and Evolution of Young Stellar Clusters}}

\author {\textbf{\large Lori Allen, S. Thomas Megeath, Robert Gutermuth}}
\affil{\small\em Harvard-Smithsonian Center for Astrophysics}

\author {\textbf{\large Philip C. Myers, Scott Wolk}} 
\affil{\small\em Harvard-Smithsonian Center for Astrophysics}

\author {\textbf{\large Fred C. Adams}} 
\affil{\small\em University of Michigan}

\author {\textbf{\large James Muzerolle, Erick Young}} 
\affil{\small\em University of Arizona}

\author {\textbf{\large Judith L. Pipher}} 
\affil{\small\em University of Rochester} 

\begin{abstract}
\baselineskip = 11pt
\leftskip = 0.65in
\rightskip = 0.65in
\parindent=1pc
{\small 
We examine the properties of embedded clusters within 1 kiloparsec 
using new data from the {\it Spitzer Space Telescope}, as well as 
recent results from 2MASS and other ground-based near-infrared surveys.
We use surveys of entire molecular clouds to understand the range and 
distribution of cluster membership, size and surface 
density. The {\em Spitzer} data demonstrate clearly that there is a continuum 
of star--forming environments, from relative isolation to dense clusters. 
The number of members of a cluster is correlated with the cluster radius,  
such that the average surface density of clusters having a few to a thousand members 
varies by a factor of only a few. 
The spatial distributions of {\it Spitzer}--identified young stellar objects  
frequently show elongation, low density halos, and sub-clustering.  The spatial 
distributions of protostars resemble the distribution of dense molecular 
gas, suggesting that their morphologies result directly from the fragmentation of the 
natal gas. 
We also examine the effects of the cluster environments on star and planet 
formation. Although Far--UV and Extreme--UV radiation from massive stars can truncate 
disks in a few million years, fewer than half of the young stars in our sample (embedded 
clusters within 1 kpc) are found in regions of strong FUV and EUV fields. Typical volume densities 
and lifetimes of the observed clusters suggest that dynamical interactions are not an important 
mechanism for truncating disks on solar system size scales.   \\~\\~\\~}%
\end{abstract}  

\section{\textbf{INTRODUCTION}}

Since PP~IV, there have been significant advances in observations of young 
stellar clusters from X-ray to millimeter wavelengths.  
But while much of the recent work has concentrated on 
the stellar initial mass function (IMF) or protoplanetary disk evolution (e.g., {\it Lada and Lada}, 2003), 
less attention 
has been directed to discerning the structure of young embedded clusters, and the evolution 
of that structure during the first few million years.  Physical properties of young embedded 
clusters, such as their shapes, sizes, and densities, should inform theories of cluster 
formation.  In this contribution, we describe recent results in which these properties are 
obtained for a representative sample of young (1-3 Myr), nearby (d$\,\le\,$1 kpc), embedded clusters.  

This contribution is motivated by three recent surveys made with 
the {\it Spitzer Space Telescope}:   
the {\it Spitzer} Young Stellar Cluster Survey -- which includes {\it Spitzer}, 
near-IR, and millimeter-wave images of 30 clusters, the {\it Spitzer} Orion 
Molecular Cloud Survey -- which covers 6.8 sq. degrees in Orion, and the Cores to Disks 
(c2d) Legacy program, which surveyed several nearby molecular clouds ({\it Evans et al.}, 2003). 
These surveys 
provide a comprehensive census of
nearly all the known embedded clusters in the nearest kiloparsec,
ranging from small groups of several stars to rich clusters with
several hundred stars.  
A new archival survey from {\it Chandra} (ANCHORS) is providing X-ray data for many 
of the nearby clusters. Since PP~IV, the {\it Two Micron All Sky Survey} (2MASS) has 
become widely used as an effective tool for  
mapping large regions of star formation, particularly in the nearby molecular clouds. 
This combination of X-ray, near-IR and mid-IR data is a powerful means for studying 
embedded populations of pre-main sequence stars and protostars. 

Any study of embedded clusters requires some method of identifying cluster members, and 
we begin by briefly reviewing methods which have progressed rapidly since PP~IV, including   
work from X-ray to submillimeter wavelengths, but with an emphasis on the mid-infrared spectrum 
covered by {\it Spitzer}. Beyond Section 2 we focus 
almost entirely on recent results from {\it Spitzer}, 
rather than a review of the literature. 
In Section 3, we discuss the cluster properties derived from large-scale surveys of young 
embedded clusters in nearby molecular clouds, including their sizes, spatial distributions, 
surface densities, and morphologies.   In Section 4 we consider the evolution of young embedded 
clusters as the surrounding molecular gas begins to disperse.   In Section 5 we discuss 
theories of embedded cluster evolution, and in Section 6 consider the impact of the cluster 
environment on star and planet formation.  Our conclusions are presented in Section 7. 

\section{\textbf{METHODS OF IDENTIFYING YOUNG STARS IN CLUSTERS}} 

\subsection{Near- and Mid-infrared}

Young stellar objects (YSOs) can be identified and classified on the basis of their
mid-infrared properties  
({\it Adams et al.}, 1987; {\it Wilking et al.}, 1989; {\it Myers and Ladd}, 1993).  
Here we review recent
work on cluster identification and characterization   
based primarily on data from the {\it Spitzer} Space Telescope.

{\it Megeath et al.} (2004) and {\it Allen et al.} (2004) developed YSO classification schemes based on
color-color diagrams from observations taken with the Infrared Array Camera
(IRAC) on {\it Spitzer}.  Examining models of
protostellar envelopes and circumstellar disks with a wide range of
plausible parameters, they found that the two types of objects
should occupy relatively distinct regions of the diagram.
Almost all of the Class I (star$+$disk$+$envelope) models exhibited the reddest colors,
not surprisingly, with the envelope density and
central source luminosity having the most significant effect
on the range of colors.  The Class II (star$+$disk) models included a treatment of
the inner disk wall at the dust sublimation radius, which is a significant
contributor to the flux in the IRAC bands.  Models of the two classes
generally occupy distinct regions in color space, indicating that they can
be identified fairly accurately from IRAC data even in the absence of
other information such as spectra.  

Comparison of these loci with YSOs
of known types in the Taurus star forming region shows reasonably good
agreement ({\it Hartmann et al.}, 2005).  Some degeneracy in the IRAC color space 
does exist; Class I sources with low envelope column densities, low mass infall 
rates or certain orientations may have the colors of Class II objects.  
The most significant source of degeneracy
is from extreme reddening due to high extinction, which can cause Class II objects to appear as
low-luminosity Class I objects when considering wavelengths
$\lambda \lesssim 10$ {\micron}.  

The addition of data from the 24 {\micron} channel of the Multiband
Imaging Photometer for {\it Spitzer} (MIPS) provides a longer wavelength
baseline for classification, particularly useful for resolving reddening
degeneracy between Class I and II.  It is also crucial for robust
identification of evolved disks, both ``transition" and ``debris",
which lack excess emission at shorter wavelengths due to the absence of
dust close to the star.  Such 24 {\micron} observations are limited, however,
by lower sensitivity and spatial resolution compared to IRAC,
as well as the generally higher background emission seen in most
embedded regions.  {\it Muzerolle et al.} (2004) delineated Class I and II
loci in an IRAC/MIPS color-color diagram of one young cluster
based on the 3.6-24 {\micron} spectral slope.  

The choice of classification method depends partly on the available data; 
not all sources are detected (or observed) in the 2MASS, IRAC, and MIPS bands. 
IRAC itself is significantly more sensitive at 3.5 and 4.5 $\mu$m than at 5.8 and 8 $\mu$m, 
so many sources may have IRAC detections in only the two shorter wavelengths, and 
require a detection in one or more near-IR bands to classify young stars 
({\it Gutermuth et al.}, 2004; {\it Megeath et al.}, 2005; {\it Allen et al.}, 2005).
{\it Gutermuth et al.} (2006) refined the IRAC+near-IR approach by correcting
for the effects of extinction, estimated from the $H-K$ color, and 
developed new classification criteria based on the extinction-corrected colors.

\begin{figure*}
\epsscale{2.0}
\plotone{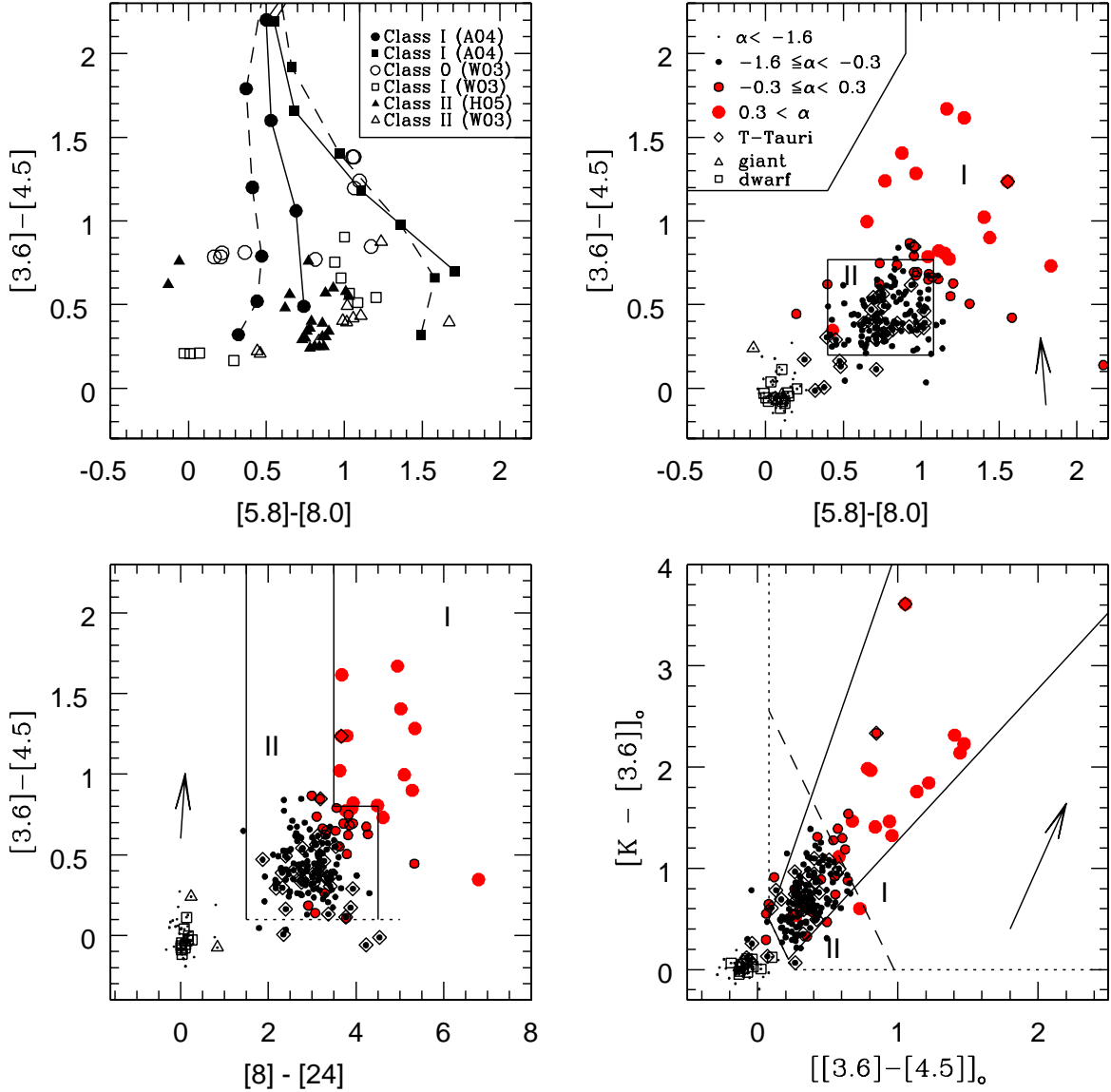}
\caption{ 
Identifying and classifying young stars using near- and mid-infrared measurements. 
In the panel at top left, a comparison of predicted IRAC colors from {\it Allen et al.} (2004) (A04), 
{\it Hartmann et al.}, (2005) (H05) and {\it Whitney et al.}, (2003) (W03). 
Triangles represent Class II models  
with ${\rm T_{eff} = 4000K}$ and a range of accretion rates, grain size distributions, 
and inclinations.  Squares and circles are Class I/0 models for a range of envelope density, 
centrifugal radius, and central source luminosity. In the remaining panels, we plot the 
data for the embedded cluster NGC2068/71 ({\it Muzerolle et al.}, 2006).  Point types are coded according to the measured 
SED slope between 2 and 24 $\mu$m.  Spectroscopically confirmed T-Tauri, giant, and dwarf 
stars are indicated. 
In the top right panel, the large rectangle marks the adopted domain of Class II sources; 
the Class I domain is above and to the right (adapted from {\it Allen et al.}, 2004). 
In the bottom right panel ({\it Gutermuth et al.}, 2006), dereddened colors are separated into 
Class I and II domains by the dashed line.  Diagonal lines outline the region where most of the 
classifiable sources are found. In the bottom left panel, the approximate domains of Class I and 
II sources are indicated by the solid lines.  The dotted line represents the adopted threshold 
for excess emission at 3.6 and 4.5 {\micron};
sources below this that exhibit large [8]-[24] excess are probably
disks with large optically thin or evacuated holes (adapted from {\it Muzerolle et al.}, 2004).  
Arrows show extinction vectors for 
${\rm A_V}=30$ ({\it Flaherty et al.}, 2006). These figures show that the various color planes considered 
here yield similar results when used to classify {\it Spitzer} sources. 
\label{newfig1}}
\end{figure*}

It is useful to compare some of the different classification schemes. 
In Fig.~\ref{newfig1} we plot first 
a comparison of the IRAC model colors from {\it Allen et al.} (2004), {\it Hartmann et al.} (2005) 
and {\it Whitney et al.} (2003).  In general, the models predict a similar range of IRAC colors 
for both Class I and Class II sources.  Also in Fig.~\ref{newfig1} we plot the same sample 
of IRAC data (NGC~2068/71) from {\it Muzerolle et al.} (2006) in three color-color planes which 
correspond to the classification methods discussed above.  In all diagrams, only those 
sources with detections in the three 2MASS bands, the four IRAC bands, and the MIPS 24 $\mu$m 
band were included.  For the sake of comparison with pre-{\it Spitzer} work, the points 
are coded according to their $K$-24$\mu$m SED slope. Prior to {\it Spitzer}, a commonly used   
4-class system was 
determined by the 2-10 $\mu$m (or 2-20 $\mu$m) slope ($\alpha$), in which 
 $\alpha > 0.3$ = Class I, $-0.3 \le \alpha < 0.3$ = ``flat'' spectrum, 
$-1.6 \le \alpha < -0.3$ = Class II, and $\alpha < -1.6$ = Class III (photosphere) 
({\it Greene et al.}, 1994).
A few of the sources in Fig. 1 have been observed spectroscopically and determined to be 
T-Tauri stars, background giants, or dwarfs unassociated with the cluster.  These are indicated.  
The diagrams also show the adopted regions of color space 
used to roughly distinguish between Class I and Class II objects.  

Classifications made with these methods are in general agreement  
with each other, though some differences are apparent. 
For example, roughly 30\% of Class I objects identified
with the {\it Allen et al.} method and detected at 24 {\micron} appear as
Class II objects in the IRAC/MIPS-24 color space, however  
many of these are borderline ``flat spectrum" sources where
the separation between Class I and II is somewhat arbitrary and may not be
physically meaningful.

These classification methods implicitly assume that all objects that
exhibit infrared excess are YSOs.  However, there can be contamination
from other sources, including evolved stars, AGN, quasars, and high-redshift
dusty galaxies.  
Since most of these unrelated objects are faint high-redshift
AGN ({\it Stern et al.}, 2005), we have found that a magnitude cut of $m_{3.6} < 14$ will remove
all but approximately 10 non-YSOs per square degree 
within each of the IRAC-only Class I and Class II loci,
and all but a few non-YSOs per square degree from the IRAC/MIPS-24 loci,  
while retaining most if not all of the cluster population.

\subsection{Submillimeter and Millimeter} 

The youngest sources in star forming regions are characterized by strong emission in the 
sub-millimeter and
far-infrared, but ususally weak emission shortward of $24$~$\mu$m.   
These  
``Class 0" objects were first discovered in sub-mm surveys of molecular 
clouds ({\it Andr\'e et al.}, 1993).
They are defined as protostars with half or more of their mass still in their envelopes, 
and emitting at least 0.5\% of their luminosity at submillimeter wavelengths. 
Motivated in part by the discovery of Class 0 objects, observers have imaged 
many embedded clusters in  
their dust continuum emission at 
millimeter and submillimeter wavelengths, revealing complex filamentary 
structure and many previously unknown sources (e.g., {\it Nutter et al.},  
2005; {\it Sandell and Knee}, 2001; {\it Motte et al.}, 2001, 1998).  

\begin{figure}[t]
\epsscale{1}
\plotone{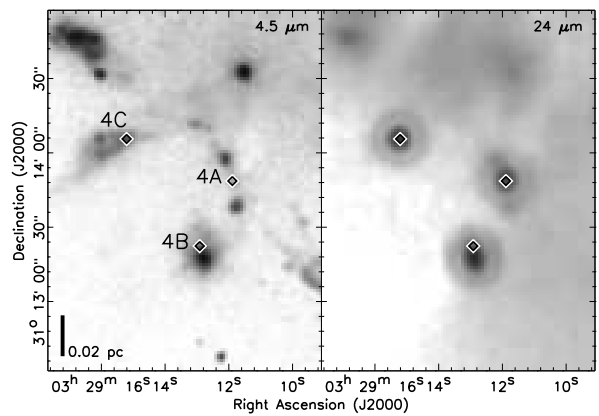}
\caption{
IRAC 4.5 $\mu$m and MIPS 24 $\mu$m images of IRAS-4 in NGC~1333. MIPS detects 
each of the three 
VLA sources, while IRAC detects their outflows but not the driving sources. 
\label{iras4abc}}
\end{figure}

These submillimeter and millimeter wavelength images generally have tens to hundreds of 
local maxima, but only a small fraction of these are ``protostars" having an internal 
heating source; the rest are ``starless cores" having a maximum of column density 
but no internal heating source. The standard way to determine whether a submm 
source is a protostar or a starless core is to search for coincidence with a 
infrared point source, such as a {\it Spitzer} source at 24 or 70 $\mu$m, or a radio continuum 
point source, such as a VLA source at 6 cm wavelength. For example the protostars 
NGC1333-IRAS 4A, 4B, and 4C in Fig.~\ref{iras4abc} are each detected at 850 $\mu$m 
({\it Sandell and Knee}, 2001), 
and each has a  counterpart in VLA observations ({\it Rodriguez et al.}, 1999) and in 
24 $\mu$m {\it Spitzer} observations, but not in the IRAC bands.  In a few cases, Class 0 protostars such as VLA1623 
have been identified from their submm emission and their radio continuum, but not from 
their mid-infrared emission, because their mid-infrared emission is too heavily 
extinguished ({\it Andr\'e  et al.}, 2000).

\subsection{X-ray}

Elevated X-ray emission is another signature of youth: young stellar
objects have typical X-ray luminosity 1000 times that of the Sun. The
e-folding decay time for this X-ray luminosity is a few 100 million
years (see e.g., {\it Micela et al.}, 1985; {\it Walter and Barry},
1991; {\it Dorren et al.}, 1995; {\it Feigelson and Montmerle}, 1999).
Although the X-ray data of young stellar clusters will be contaminated
by AGN and other sources, this contamination can be reduced by
identifying optical/infrared counterparts to the X-ray sources. X-ray
sources where the ratio of the X-ray luminosity to the bolometric
luminosity (L$_X / {\rm L_{bol}}$) ranges from  0.1\% to 0.01\% are
likely pre-main sequence stars.  In contrast to the infrared
techniques described in 2.1, which can only identify Class 0/I and II
sources; X-ray observations can readily detect class II and class III
objects, with perhaps some bias toward class III objects ({\it
  Flacomio et al.}, 2003).  The main limitation of X-ray observations
is the lack of sensitivity toward lower mass stars.  A complete sample
of stars requires a sensitivity toward souces with luminosities as low
as 10$^{27}$ erg cm$^{-2}$~s$^{-1}$ ({\it Feigelson et al.}, 2005),
the sensitivity of most existing observations are an order of
magnitude higher. The observed X-ray luminosity is also affected by
extinction. Depending on the energy of the source, the sensitivity can
be reduced by a factor of ten for sources at A$_V\sim$10 ({\it Wolk et  
 al.}, 2006).

\subsection{Emission Lines and Variability} 

Among other techniques for identifying young cluster members, spectroscopic surveys 
for emission lines and photometric surveys for variability have been used successfully 
at visible and near-IR wavelengths.  
The most common means of identifying young stars spectroscopically is through 
detection of optical emission lines, in particular H$\alpha$ at 6563 \AA ~({\it Herbig and Bell}, 1988). 
Large-scale objective prism ({\it Wiramihardja et al.}, 1989; {\it Wilking et al.}, 1987) and later, 
wide-field multi-object spectroscopy (e.g., {\it Hillenbrand et al.}, 1993) has been 
effective in identifying young stars in clusters and throughout molecular clouds, however they 
miss the deeply embedded members that are optically faint or invisible. 
This problem is partly alleviated by large-scale surveys for photometric variability in the optical and near-IR. 
Recent near-IR surveys by {\it Kaas} (1999) and {\it Carpenter et al.} (2001, 2002) have been 
successful at identifying 
young cluster members in Serpens, Orion and Chamaeleon, respectively.  

\subsection{Star Counts} 

Much of the work on the density, size, and structure of embedded
clusters has relied on using star counts; indeed, the distribution of
2.2~$\mu$m sources were used to identify clusters in the Orion B cloud
in the seminal work of {\it Lada et al.} (1991).  Instead of identifying
individual stars as members, methods based on star counts include all
detected sources and employ a statistical approach toward membership, in
which an average density of background stars is typically estimated
and subtracted out.  In this analysis, the star counts are typically
smoothed to produce surface density maps; a variety of smoothing
algorithms are in the literature ({\it Gomez et al.}, 1993; {\it Gladwin et al.}, 1999;
{\it Carpenter}, 2000; {\it Gutermuth et al.}, 2005; {\it Cambresy et al.}, 2006)

The degree of contamination by foreground and background stars is the
most significant limitation for star count methods, and the efficacy of
using star counts depends strongly on the surface density of contaminating
stars.  In many cases, the contamination can be minimized by setting a
$K$--band brightness limit ({\it Gutermuth}, 2005; {\it Lada et
al.}, 1991).  To estimate the position dependent contamination by field
stars, models or measurements of the field star density can be
combined with extinction maps of the molecular cloud ({\it Carpenter}, 2000;
{\it Gutermuth et al.}, 2005; {\it Cambresy}, 2006).  These maps are subtracted from
the surface density of observed sources to produce maps of the
distribution of embedded stars; however, these maps are still limited by
the remaining Poisson noise from the subtracted stars.

\begin{figure*}[t]
\epsscale{1.7}
\plotone{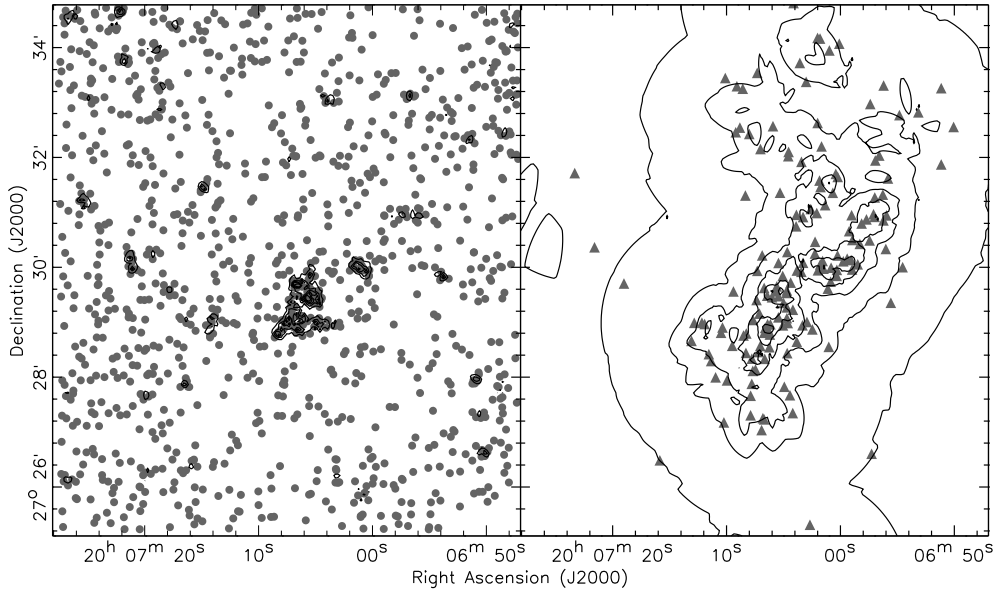}
\caption{IRAS~20050 surface densities derived from the statistical technique applied to all stars (left),
and from identifying the infrared-excess sources (right).  In the left panel, all sources having K$<$16 
are plotted as a function of their position. Contours show the surface density of K-band sources, starting 
at 1450 pc$^{-2}$ (5$\sigma$ above median field star density) and increasing at intervals of 750 pc$^{-2}$.
In the right panel, sources with infrared excess emission are plotted, and contours of their surface 
density are shown for 10, 60, 160, 360, 760, and 1560 pc$^{-2}$.  The statistical technique (left) yields   
a higher peak surface density ($\sim$6000 pc$^{-2}$ at the center) than the IR-excess 
technique ($\sim$3000 pc$^{-2}$), but the latter is more sensitive to the spatially extended population 
of young stars.  
\label{fig:irasdens}}
\end{figure*}

Star count methods have the advantage that they do not discriminate
against sources without infrared excess, bright X-ray emission,
variability, or some other indication of youth.  On the other hand,
they only work in regions where the surface density of member stars is
higher than the statistical noise from contaminating field stars.  In
Fig.~\ref{fig:irasdens} we show maps of the IRAS~20050 cluster derived
from the $K$-band star
counts and from the distribution of infrared-excess sources.  In the
case of IRAS~20050, we find that the star count method provides a
better map of the densest regions (due in part to confusion with
bright nebulosity and sources in the {\it Spitzer} data), while the lower
density regions surrounding these peaks are seen only in the
distribution of {\it Spitzer} identified infrared excess sources (due to the
high density of background stars).

\section{THE STRUCTURE AND EVOLUTION OF CLUSTERS: OBSERVATIONS}

\subsection{Identifying Clusters in Large Scale Surveys of Molecular Clouds}

Unlike gravitationally bound open clusters or globular clusters,
embedded clusters are not isolated objects. In most cases, molecular
cloud complexes contain multiple embedded clusters as well as
distributed populations of relatively isolated stars.  Recent large
scale surveys and all sky catalogs are now providing new opportunities
to study the properties of embedded clusters through surveys of entire
molecular clouds.  The advantage of studying clusters by surveying
entire molecular clouds is twofold.  First, the surveys provide an
unbiased sample of both the distributed and clustered populations
within a molecular cloud.  Second, the surveys result in an unbiased
measurement of the distribution of cluster properties within a single
cloud or ensemble of clouds.  For the remainder of this discussion, we
will use the word ``cluster" to denote embedded clusters of young
stars. Most of these clusters will not form bound open
clusters ({\it Lada and Lada}, 1995).

We now concentrate on two recent surveys for young stars in relatively
nearby ($<$1 kpc) molecular clouds.  {\it Carpenter} (2000) used the
2MASS 2nd incremental point source catalog to study the distribution
of young stars in the Orion A, Orion B, Perseus and Monoceros R2
clouds. Since the 2nd incremental release did not cover the entire sky,
only parts of the Orion B and Perseus clouds were studied. More
recently, {\it Spitzer} has surveyed a number of molecular clouds. We
discuss here new results from the {\it Spitzer} Orion Molecular Cloud Survey
({\it Megeath et al.}, 2006) and the Cores to Disks (c2d) Legacy program
survey of the Ophiuchus Cloud ({\it Allen et al.}, 2006). We use these
data to study the distribution of the number of cluster 
members, the cluster radius, and the stellar density in this small sample of
clouds.

The advantage of using these two surveys is that they draw from
different techniques to identify populations of young stellar
objects. The analysis of the 2MASS data relies on star counting
methods (Section 2.5), while the {\it Spitzer} analysis relies on
identifying young stars with infrared-excesses from combined {\it Spitzer} 
and 2MASS photometry (Section 2.1; {\it Megeath et al.} 2006).  The
2MASS analysis is limited by the systematic and random noise from the
background star subtraction, making the identification of small groups
and distributed stars subject to large uncertainties.  The {\it Spitzer}
analysis is limited to young stars with disks or envelopes.  A
significant number of young stars in embedded clusters do not show
excesses; this fraction may range from 20\% to as much as 50\% for
1-3~Myr clusters ({\it Haisch et al.}, 2001).

\begin{figure*}
\epsscale{2.0} 
\plotone{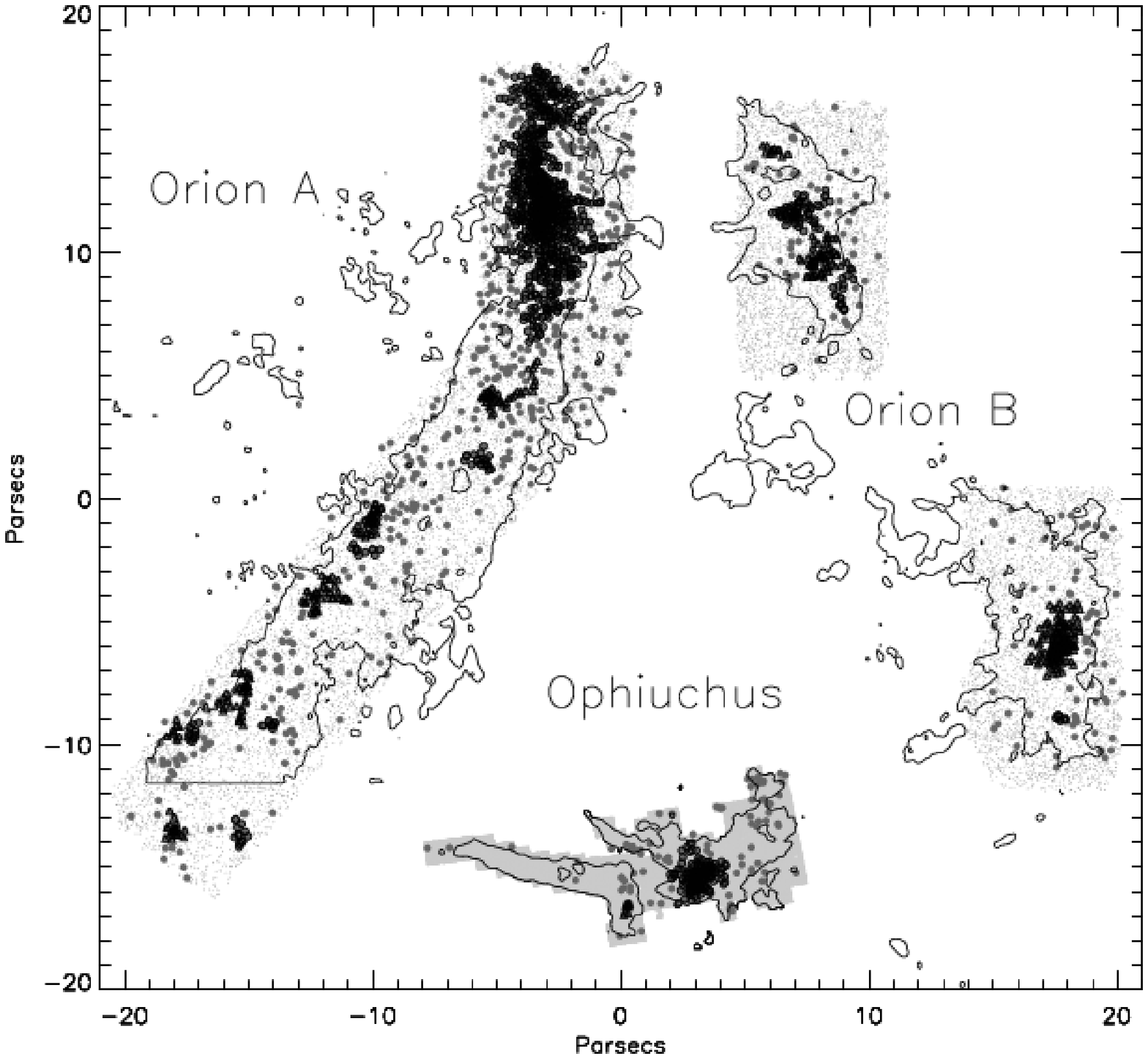}
\caption{
The spatial distribution of all Spitzer identified infrared excess sources from the combined 
IRAC and 2MASS photometry of Orion A (left), Orion B (right) and Ophiuchus (bottom center).  
The contours outline the Bell Labs $^{13}$CO maps for the Orion A and B clouds ({\it Bally et al.}, 
1987; {\it Miesch and Bally}, 1994),  and an $A_V$ map of Ophiuchus ({\it Huard}, 2006).  
The small grey dots show all the detections in the Spitzer 3.6 and 4.5~$\mu$m bands with magnitudes 
brighter than 15th and uncertainties less than 0.15.  The large grey dots are the
sources with infrared excesses.  The black circles and triangles are sources found in 
clusters using the method described in Section 3.1; the two symbols are alternated so that
neighboring clusters can be differentiated.  
Note that there are two clusters in the Orion A cloud which are below the lower 
boundary of the Bell Labs map. Each of the clouds has a significant distributed population 
of IR-excess sources.  
\label{fig:spitzercluster}}
\end{figure*}

{\it Carpenter} (2000) identified stellar density peaks 
more than six times the RMS background noise, and
defined a cluster as all stars in a closed 2$\sigma$ contour surrounding these
peaks. 
{\it Megeath et al.} (2006) defined clusters as groups of 10 or more IR-excess
sources in which each member is within a projected distance of 0.32~pc
of another member (corresponding to a density of 10 stars pc$^{-2}$).
Only groups of ten or more neighbors are considered clusters.  The
clusters identified in the {\it Spitzer} survey are shown in
Fig.~\ref{fig:spitzercluster}.

\subsection{The Fraction of Stars in Large Clusters}

It is now generally accepted that most stars form in clusters ({\it Lada and Lada}, 1995), but
quantitative estimates of the fraction of stars which form in large
clusters, small clusters, groups and relative isolation are still
uncertain.  {\it Porras et al.} (2003) compiled a list of all known
groups and clusters with more than 5 members within 1 kpc of the Sun,
while {\it Lada and Lada} (2003) compiled the properties of a sample of
76 clusters with more than 36 members within 2 kpc.  Although these
compilations are not complete, they probably give a representative
sample of clusters in the nearest 1-2 kpc. In the sample of {\it
  Porras et al.} (2003), 80\% of the stars are found in clusters with
$N_{star} \ge 100$, and the more numerous groups and small clusters 
contain only a small fraction of the stars (also see
{\it Lada and Lada}, 2003).

\begin{figure}[t]
\epsscale{0.9} 
\plotone{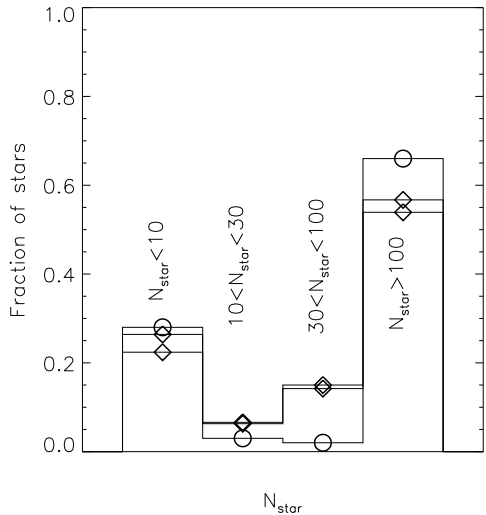}
\caption{The distribution of the fraction of stars in clusters taken from
{\it Carpenter} (2000) (circles) and the {\it Spitzer} surveys of Orion and
Ophiuchus (diamonds).  The {\it Spitzer} surveys show a range, depending on 
whether corrections are made for background AGN.  In both the 2MASS and 
{\it Spitzer} surveys, the distributed population 
($N_{star} < 10$) accounts for more than 20\% of the total number of stars. 
\label{fig:cumulative}}
\end{figure}

In Fig.~\ref{fig:cumulative}, we plot the fraction of members from
the 2MASS and {\it Spitzer} surveys as a function of the number of cluster
size.  Following the work of {\it Porras et al.} (2003), we divide the
distribution into four sizes: $N_{star} \ge 100$, $100 > N_{star}
\ge 30$, $30 > N_{star} \ge 10$, and $N_{star} < 10$.  The main
difference from the previous work is that we include a bin for
$N_{star} < 10$; these we refer to as the distributed population.
All of the observed molecular clouds appear to contain a distributed
population.  {\it Carpenter} (2000) estimated that the fraction of stars in
the distributed population were 0\%, 20\%, 27\%, and 44\% for the
Orion B, Perseus, Orion A and Mon R2 cloud, respectively, although the
estimated fraction ranged from 0-66\%, 13-41\%, 0-61\% and 26-59\%,
depending on the assumptions made in the background star subtraction.
In the combined {\it Spitzer} survey sample, the fraction of distributed
stars is 32-11\%, 26-24\%, and 25-21\% for the Ophiuchus, Orion A and Orion
B clouds respectively.  The uncertainty is due to contamination from
AGN: the higher fraction assumes no contamination, the lower number
assumes that the distributed population contains 10 AGN for every
square degree of map size.  The actual value will be in between those;
extinction from the cloud will lower the density of AGN, and some of
the contaminating AGN will be found toward clusters.  In total, these
measurements suggest that typically 20-25\% of the stars are in the
distributed population.

There are several caveats with this analysis.  The first is the lack
of completeness in the existing surveys.  {\it Carpenter} (2000)
considered the values of $N_{star}$ as lower limits due to
incompleteness and due to the masking of parts of the clusters to
avoid artifacts from bright sources.  Completeness is also an issue in
the center of the Orion Nebula Cluster (ONC) for the {\it Spitzer} 
measurements.  
Also, we have not
corrected the {\it Spitzer} data for the fraction of stars which do not show
infrared excesses, the actual number of stars may be as much as a
factor of two higher ({\it Gutermuth et al.}, 2004).

Another uncertainty is in the definition of the clusters.  
The clusters identified by these two methods are not entirely
consistent.  For example, in Orion A there is an uncertainty in the
boundaries of the ONC.  There is a large halo of stars surrounding this
cluster, and the fraction of young stars in large clusters is dependent on
whether stars are grouped in the ONC, in nearby smaller groups, or the distributed population. 
Both the 2MASS and the {\it Spitzer} data lead to an expansive definition of
this cluster, extending beyond the Orion Nebula and incorporating the
OMC2/3 and NGC 1977 regions, as well the L1641 North group for the
2MASS analysis.  The resulting cluster contains a significant number
of stars in a relatively low stellar density environment far from the
O-stars exciting in the nebula, which differs significantly from the
environment of the dense core of the cluster embedded in the Orion
Nebula.  The treatment of the ONC is critical to this analisys: 50\%
(for the 2MASS sample) to 76\% (for the {\it Spitzer} sample) of the stars in
large clusters ($N_{star} \ge 100$) are found in the ONC.

A final caveat is that these results apply to the current epoch of
star formation in the nearest kiloparsec. While the largest cluster within
1 kpc is the ONC with 1000-2000 members, a growing
number of young super star clusters, which contain many thousands of
stars, have been detected in our Galaxy.  Super star clusters may
bridge the gap between embedded clusters in then nearest kiloparsec,
and the progenitors of the globular clusters which formed earlier in
our Galaxy's history.  Thus, the distribution of cluster sizes we have
derived may not be representative for other regions of the Galaxy, or
early epochs in our Galaxy's evolution.

\subsection{The Surface Density of Stars in Embedded Clusters}

\begin{figure}[t]
\epsscale{0.9} 
\plotone{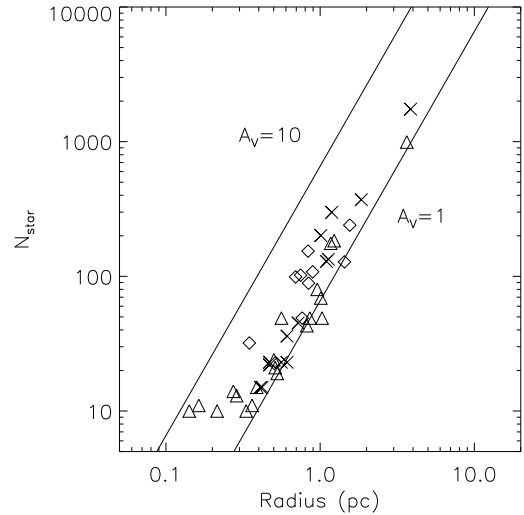} 
\caption{$N_{star}$ {\it vs.} cluster radius for the 
  2MASS survey (crosses) of {\it Carpenter} (2000), the {\it Spitzer} Ophiuchus
  and Orion surveys (triangles) of {\it Megeath et al.} (2006) and {\it Allen et al.} (2006), and the 
{\it Spitzer} 
  young stellar cluster survey (diamonds).  Lines of constant column density
  are shown for a column density for ${\rm A_V=1}$ and ${\rm A_V=10}$. The average surface density 
of cluster members varies by less than an order of magnitude. 
\label{fig:radiussize}}
\end{figure}

In a recent paper, {\it Adams et al.} (2006) found a correlation between the
number of stars in a cluster and the radius of the cluster, 
using the tabulated cluster properties in
{\it Lada and Lada} (2003). They found that the correlation is even stronger
if only the 2MASS identified clusters from {\it Carpenter} (2000) were used, 
in which case the parameters were derived in a
uniform manner.   
The same correlation is
seen in a sample of clusters defined by {\it Spitzer} identified IR-excess 
sources.  This correlation is shown for the 2MASS and {\it Spitzer} samples
in Fig.~\ref{fig:radiussize}.  This relationship shows that while
$N_{star}$ varies over 2 orders of magnitude
and the cluster radius ($R_{cluster}$) varies by almost 2 orders of
magnitude, the average surface density of cluster members ($N_{star}/\pi
R_{cluster}^2$) varies by less than one order of magnitude.  
The lower surface (${\rm A_V = 1}$) envelope of this
correlation may result in part from the methods used to identify  
clusters.  In particular, for the many clusters surrounded by large, low surface density halos of stars, the
measured radius and density of these clusters depends on the threshold surface density or spatial separation
used to distinguish the cluster stars from those in the halos.
We can  
convert the surface densities of members into column densities of mass
by assuming an average stellar mass of $0.5$~M$_\odot$.  Assuming a
standard abundance of hydrogen, and the typical conversion from
hydrogen column density to ${\rm A_V}$, we plot lines of constant ${\rm A_V}$ in
Fig.~\ref{fig:radiussize}.  In this figure the clusters are
bracketed by lines equivalent to ${\rm A_V \sim 1}$ and ${\rm A_V \sim 10}$.
Interestingly, this result is similar to one of Larson's laws for
molecular clouds, that the average column density of gas in molecular
clouds is independent of cloud size and mass ({\it Larson}, 1985; see also
the chapter by {\it Blitz et al.}). 

\subsection{The Spatial Structure of Embedded Clusters}

\begin{figure*}[t]
\epsscale{1.8}
\plotone{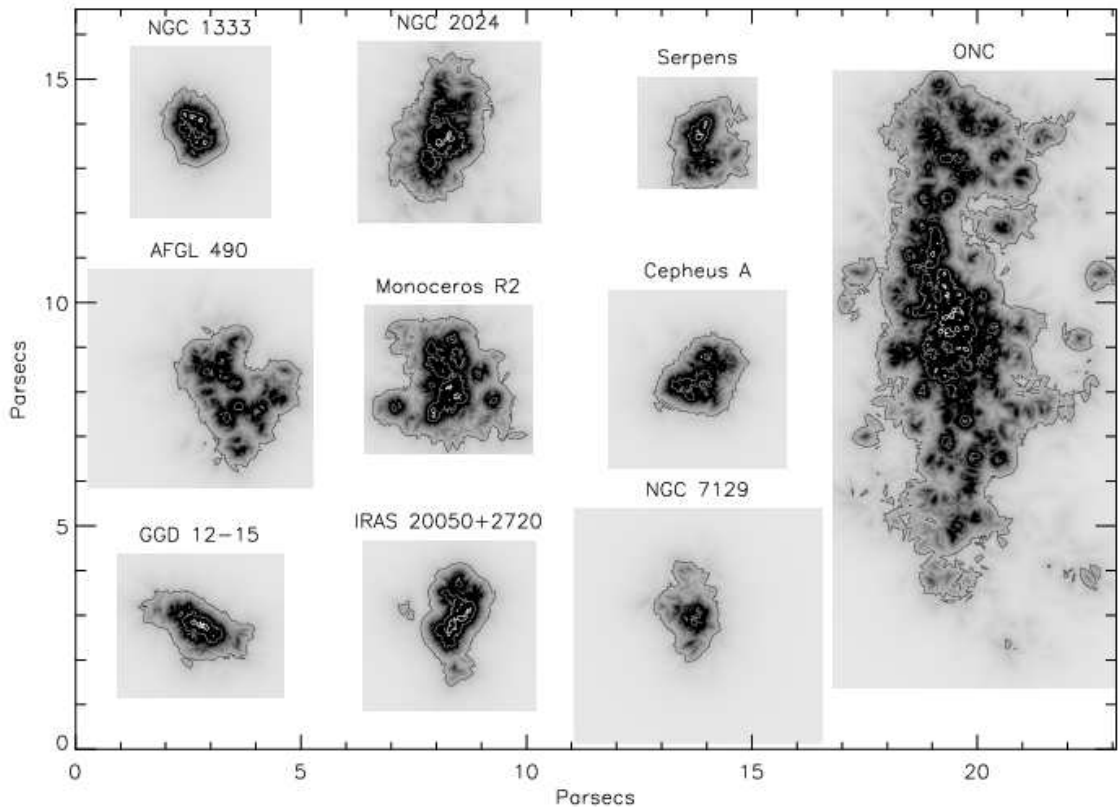}
\caption{The distribution of infrared excess sources in ten clusters 
  surveyed with {\it Spitzer}.  The contours are
  at 1, 10 and 100 IR-excess sources pc$^{-2}$. These data clearly show
  that clusters are not circularly symmetric, but are often elongated.
  Some of the clusters, such as IRAS~20050, show distinct clumpy
  structure, although much of the small scale structure seen in the
  highest contours is due to statistical fluctuations in the smoothing
  scale.  The three most circularly symmetric clusters are Cepehus A,
  AFGL 490 and NGC 7129; the irregular structure in these clusters is
  due in part to statistical fluctuations in regions of lower
  surface density.
\label{fig:album}}
\end{figure*}

One of the major
goals of the {\it Spitzer} young stellar cluster and Orion surveys is to
systematically survey the range of cluster morphologies by identifying
the young stellar objects with disks and envelopes in these clusters.
An initial result of this effort is displayed for ten clusters in
Fig.~\ref{fig:album}, which shows the surface density of IR-excess 
sources.  In this section, we give a brief overview of the common structures found
in embedded clusters, both in the literature and in the sample of
clusters imaged with {\it Spitzer}.  We also discuss {\it ISO} and {\it Spitzer}
observations of the youngest objects in these
regions, the Class I and 0 sources.

Many of the clusters shown in Fig.~\ref{fig:album}
appear elongated; this had also been evident in some of the earlier
studies of clusters ({\it Carpenter et al.}, 1997; {\it Hillenbrand and  
Hartmann}, 1998). To quantify this asymmetry, {\it Gutermuth et al.} (2005,
2006) compared the distribution of stars as a function of position angle 
to Monte Carlo simulations of circularly symmetric clusters, and 
demonstrated that the 
elongation is statistically significant in three of the six
clusters in their sample.  The elongation appears to be a result of
the primordial structure in the cloud; for the two elongated clusters
which have 850~$\mu$m dust continuum maps, the elongation of the
cluster is aligned with filamentary structure seen in the parental
molecular cloud.  This suggests that the
elongation results from the formation of the clusters in highly
elongated, or filamentary, molecular clouds.

Not all clusters are elongated. {\it Gutermuth et al.} (2005) found no
significant elongation of the NGC 7129 cluster, a region which also
showed a significantly lower mean and peak stellar surface density
than the more elongated clusters in his sample.  Since the cluster was
also centered in a cavity in the molecular cloud (see Section 5);
they proposed that the lack of elongation was due
to the expansion of the cluster following the dissipation of the
molecular gas. However not all circularly symmetric clusters are easily
explained by expansion; {\it Gutermuth et al.} (2006) find two deeply
embedded clusters with no significant elongation or clumps, but no
sign of the gas dispersal evident in NGC 7129.  These two clusters,
Cepheus A and AFGL 490, show azimuthal symmetry, which   
may reflect the primordial structure of the cluster.

Examination of Fig.~\ref{fig:album} reveals another common structure:
low density halos surrounding the dense centers, or cores, of the
clusters.  With the exception of AFGL~490 and perhaps Cepheus A, all
of the clusters in Fig.~\ref{fig:album} show cores and halos.  The
core-halo structure of clusters has been studied quantitatively
through azimuthally smoothed radial density profiles ({\it Muench et
al.} 2003).  Although these density profiles can be fit by power laws,
King models, or exponential functions ({\it Hillenbrand and Hartmann}, 1998;
{\it Lada and Lada}, 1995; {\it Horner et al.}, 1997; {\it Gutermuth}, 2005), the resulting
fits and their physical implications can be misleading.  As pointed
out by {\it Hartmann} (2004), azimuthally averaged density profiles can be
significantly steepened by elongation ({\it Hartmann} (2004) argues this for
molecular cores, but the same argument applies to clusters).  A more
sophisticated treatment is required to study the density profiles of
elongated clusters.

It has long been noted that young stellar clusters are sometimes
composed of multiple sub-clusters ({\it Lada et al.}, 1996; {\it Chen et
al.}, 1997; {\it Megeath et al.}, 1996; {\it Allen et al.}, 2002; {\it Testi}, 2002).  
Clusters with multiple density peaks
or sub--clusters were classified as heirarchical clusters by {\it Lada and 
Lada} (2003). 
In some cases it is difficult to distinguish between two
individual clusters and sub--clusters within a single cluster.  An
example are the NGC 2068 and NGC 2071 clusters in the Orion B cloud
(Fig.~\ref{fig:spitzercluster}). These appear as two peaks in a more
extended distribution of stars, although the cluster identification
method described in Section 3.1 separated the two peaks into two
neighboring clusters.  In the sample of {\it Gutermuth et al.} (2005, 2006),
clumpy structure was most apparent in the IRAS 20050 cluster (also
see {\it Chen et al.}, 1997).  In this cluster, the sub--clusters are
asociated with distinct clumps in the 850~$\mu$m map of the associated
molecular cloud.  This suggests that like elongation, sub-clusters
result from structures in the parental molecular cloud.

\subsection{The Distribution of Protostars}

\begin{figure*}[t]
\epsscale{1.5} 
\plotone{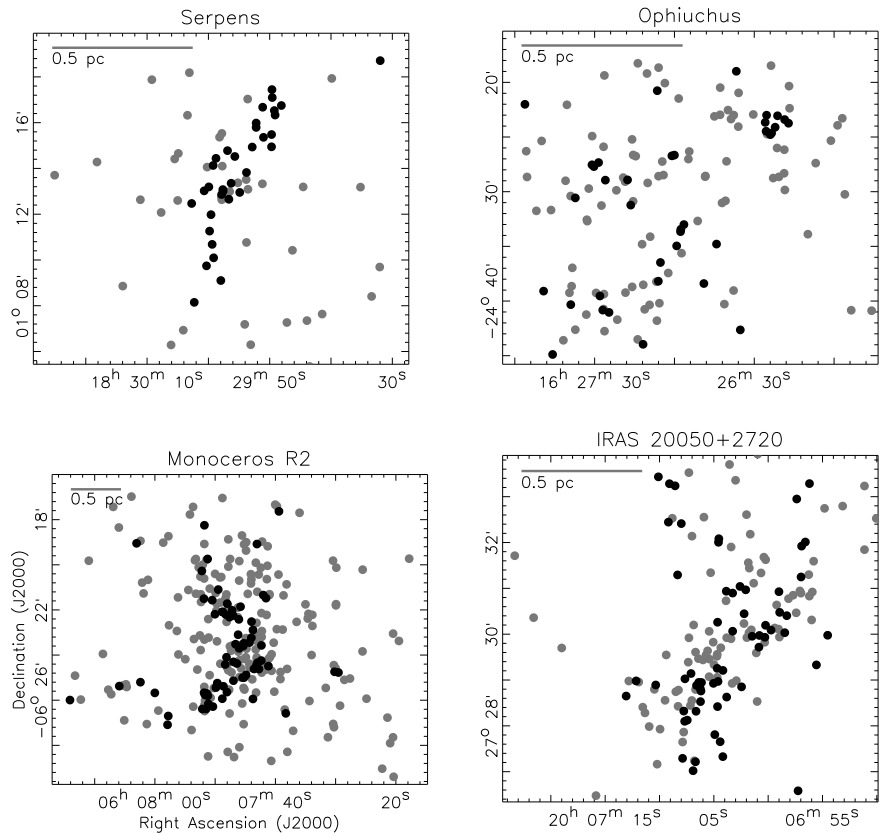}
\caption{The spatial distributions of {\it Spitzer} identified class I/0 (dark circles) 
and Class II (light circles) objects in four clusters: L1688 in Ophiuchus, Serpens, 
Mon R2 and IRAS~20050.  The Class I/0 sources are often distributed along filamentary 
structures, while the Class II sources are more widely distributed.  Many small groups 
of protostars are dense enough that interactions bewteen individual objects may occur. 
\label{fig:class1}}
\end{figure*}

If the observed morphologies of embedded clusters result from the 
filamentary and clumpy nature of the
parental molecular clouds,  
then the younger Class 0/I objects, which have had the least time to
move away from their star formation sites, should show more
pronounced structures than the older, pre-main sequence Class II and
Class III stars.
{\it Lada et al.} (2000) found a deeply embedded population of young stellar
objects with large $K-L$ colors toward the ONC; these
protostar candidates showed a much more elongated and clumpy structure
than the young pre-main sequence stars in the Orion Nebula.  Using the
methods described in Section 2.1, we have identified Class 0/I and II
objects in clusters using combined {\it Spitzer} and ground-based near-IR
photometry.  In Fig.~\ref{fig:class1}, we plot the distribution of
class 0/I and II sources for four clusters in our sample.  In the
L1688 and IRAS 20050 clusters, the protostars fall preferentially in
small sub-clusters, and are less widely distributed than the Class II
objects.  In the Serpens and GGD~12-15 clusters, the protostars are
organized into highly elongated distributions. An interesting example
containing multiple elongated distributions of protostars is the
"spokes" cluster of NGC 2264, which shows several linear chains of
protostars extending from a bright infrared source ({\it Teixeira et
al.}, 2006).  These chains, which give the impression of spokes on a
wheel, follow filamentary structures in the molecular cloud.  These
data support the view that the elongation and sub-clustering are
indeed the result of the primordial distribution of the parental dense 
gas.  It is less clear whether the observed halos result from  
dynamical evolution or originate {\it in situ} in less active regions of star formation
surrounding the more active cluster cores.  The current data
suggest that the halos are at least in part primordial; class 0/1
objects are observed in the halos of many clusters ({\it Gutermuth et
al.}, 2004; {\it Megeath et al.}, 2004).

The spacing of protostars is an important constraint on the physical
mechanisms for fragmentation and possible subsequent interactions by
protostars.  {\it Kaas et al.} (2004) analyzed the spacing of Class I and II
objects identified in {\it ISO} imaging of the Serpens cluster.  They
calculated the separations of pairs of Class I objects, and found that
the distribution of these separations peaked at 0.12 pc.  In
comparison, the distribution of separations for Class II objects show
only a broad peak at 0.2 to 0.6~pc; this reflects the more spatially
confined distributions of protostars discussed in the previous
section.  {\it Teixeira et al.} (2006) performed a similar analysis for the
sample of protostars identified in the spokes cluster of NGC 2264.  The
distribution of nearest neighbor separations for this sample peaked
at 0.085 pc; this spacing is similar to the Jeans length calculated
from observations of the surrounding molecular gas.

Although the observed typical spacing of protostars in Serpens and NGC
2264 apears to be $\sim$ 0.1 pc, as shown in Fig.~\ref{fig:class1}, dense
groups of protostars are observed in both these regions (and others)
in which several Class I/0 sources are found within a region 0.1 pc in
diameter.  This is the distance a protostar could move in   
100,000 years (the nominal protostellar lifetime) at a velocity of
${\rm 1~kms^{-1}}$. This suggests that if the velocity dispersion of
protostars is comparable to the turbulent velocity dispersion observed
in molecular clouds, interactions between protostars may occur,
particularly in dense groupings.  On the other hand, observations of some 
dense star--forming clumps show motions through their envelopes much less 
than ${\rm 1~kms^{-1}}$ ({\it Walsh et al.} 2004). The densest grouping of protostars
so far identified in the {\it Spitzer} survey is found in the spokes
cluster.  One of the protostars in the spokes has been resolved into a
small system of 10 protostars by ground-based near-IR imaging and by
{\it Spitzer} IRAC imaging. These protostars are found in a region 10,000 AU
in diameter.  It is not clear whether these objects are in a bound
system, facilitating interactions as the sources orbit within the
system, or whether the stars are drifting apart as the
molecular gas binding the region is dispersed by the evident outflows  
({\it Young et al.}, 2006).  It should be noted that this group
of 10 protostars appears to be the only such system in the spokes
cluster.  Thus, although dense groups of   
protostars are present in star forming regions, they may not be
common.

\section{GAS DISRUPTION AND THE LIFETIME OF EMBEDDED CLUSTERS}

In the current picture of cluster evolution, star formation is
terminated when the parental gas has dispersed.  An understanding of the
mechanisms and time scales for the disruption of the gas is necessary
for understanding the duration of star formation in clusters, the
lifetime and eventual fate of the clusters, and the ultimate star
formation efficiency achieved in a molecular cloud.

The most massive stars have a disproportionate effect on cluster
evolution. Massive O stars can rapidly disrupt the parental molecular
cloud through their ionizating radiation.  The effect of the
disruption is not immediate; once massive stars form in a molecular
core, star formation may continue in the cluster while the massive
star remains embedded in an ultracompact HII region.  Examples of
clusters in this state within 1~kpc of the Sun are the GGD~12-15 and
Mon~R2 clusters.  The timescale for the disruption of the core is
equivalent to the lifetime of the ultracompact HII region ({\it Megeath et
al.}, 2002); this lifetime is thought to be $\sim 10,000$ years for the
solar neighborhood ({\it Casussus et al.}, 2000; {\it Comeron and Torra}, 1996). 

In our sample of nearby embedded clusters, most systems do not contain
O-stars.  However, a number of partially embedded clusters in the
nearest 1 kpc show evidence for significant disruption by B type
stars.  Due to the partial disruption of the clouds, the
clusters in these regions are found in cavities filled with emission
from UV heated polycyclic aromatic hydrocarbons ({\it Gutermuth et
al.}, 2004). 
The time scale for the
disruption by B stars can be estimated using measurements of the ages
of the clusters.  In our survey of nearby regions, we have three
examples of regions with such cavities: NGC 7129 (earliest member B2),
IC 348 (earliest member B5) and IC5146 (earliest member B0-1), with ages
2~Myr, 3~Myr and 1~Myr, respectively ({\it Hillenbrand et al.}, 1992; {\it Hillenbrand}, 
1995; {\it Luhman et al.}, 2003; {\it Herbig and Dahm}, 2002).  
The presence of
large, UV illuminated cavities in these regions suggest that the
non-ionizing far--ultraviolet radiation (FUV) from B-stars may be
effective at heating and evaporating molecular cloud surfaces in cases
where intense FUV radiation from O-stars is not present. For example,
in the case of NGC~7129, {\it Morris et al.} (2004) find that the temperature
at the molecular cloud surface has been heated to 700~K by the FUV
radiation.  Future work is needed to determine if the high
temperatures created by the FUV radiation can lead to substantial
evaporative flows.

In regions without OB stars, however, some other mechanism must operate. An
example is IRAS 20050.  Based on SCUBA maps, as well as the reddening
of the members, {\it Gutermuth et al.} (2005) found that the cluster is
partially offset from the associated molecular gas, suggesting that
the gas had been partially dispersed by the young stars
Although this region contains no OB stars, it
displays multiple outflows ({\it Chen et al.} 1997). Another example may be
the NGC 1333 cloud, where {\it Quillen et al.} (2005) found evidence of
wind-blown cavities in the molecular gas.  In these regions, outflows
may be primarily responsible for dissipating the dense molecular gas
(e.g., {\it Matzner and McKee} 2000).

It is important to note that star formation continues during the gas
dissipation process.  Even when the gas around the main cluster has
been largely disrupted (such is the case in the ONC,
IC 348 and NGC 7129), star formation continues on the outskirts of the
cluster in regions where the gas which has not been removed. Thus, the
duration of star formation in these regions appears similar to the gas
dispersal time of $\sim 1-3$~Myr.  Older clusters have not been found
partially embedded in their molecular gas ({\it Leisawitz et al.}, 1989).

\section{\textbf{EARLY CLUSTER EVOLUTION}}

Theories of cluster formation are reviewed elsewhere in these 
proceedings (see the chapters by {\it Ballesteros-Paredes et al.}
and {\it Bonnell et al.}).  Here we will discuss the dynamical evolution of 
young clusters during the first few million years.  

Although most stars seem to form within clusters of some type (see
Section 4), only about ten percent of stars are born within star-forming
units that are destined to become open clusters.  As a result, for
perhaps 90 percent of forming stars, the destruction of their birth
aggregates is an important issue.  
Star formation in these systems is not 100 percent
efficient, so a great deal of cluster gas remains in the system. This
gaseous component leaves the system in a relatively short time (a few
Myr -- see above) and its departure acts to unbind the cluster.  At
the zeroth level of understanding, if the star formation efficiency (SFE)  
is less than 50\% , then a substantial amount of unbinding
occurs when gas is removed. However, this description is overly
simple. The stars in the system will always have a distribution of
velocities. When gas is removed, stars on the high velocity tail of
the distribution will always leave the system (even for very high
SFE) and those on the extreme low velocity tail will tend to stay
(even for low SFE). The fraction of stars that remain bound after gas
removal is thus a smooth function of star formation efficiency
(several authors have tried to calculate the function: see {\it Adams}, 2000;
{\it Boily and Kroupa}, 2003a, 2003b; {\it Lada et al.}, 1984). 
The exact form of  
the bound fraction, $f_b (\epsilon)$, which is a function of SFE,  
depends on many other cluster properties: gas removal rates,
concentation of the cluster, total depth of the cluster potential
well, the distribution functions for the stellar velocities (radial vs
isotropic), and the spatial profiles of the gaseous and stellar
components (essentially, the SFE as a function of radial position). 
At the crudest level, the bound fraction function has the form $f_b \approx 
\sqrt{\epsilon}$, but the aforementioned complications allow for a
range of forms.

The manner in which a cluster spreads out and dissolves after its gas
is removed is another important problem. After gas removal, clusters
are expected to retain some stars as described above, but such systems
are relatively short-lived.  For example, consider a cluster with $N$
= 100 in its early embedded phase, before gas removal. After the gas
leaves, typically one half to two thirds of the stars will become
unbound along with the gas.  The part of the cluster that   
remains bound will thus contain only $N$ = 30 -- 50 stars. Small groups
with $N < 36$ have relaxation times that are shorter than their
crossing times ({\it Adams}, 2000) and such small units will exhibit
different dynamical behavior than their larger counterparts. In
particular, such systems will relax quickly and will not remain
visible as clusters for very long.

As more data are taken, another mismatch between theory and
observations seems to be emerging: The theoretical calculations
described above start with an established cluster with a well-defined
velocity distribution function, and then remove the gaseous component
and follow the evolution. 
Given the constant column density relationship for clusters (section 3.4), 
that the velocity of the stars are virialized,  
and assuming that 30\% of the cluster mass is in stars 
(see {\it Lada and Lada}, 2003), then the crossing time for the typical 
cluster in our sample is $\sim 1$~Myr 
(although it can be shorter in the dense centers of clusters).
As a result, in
rough terms, the gas removal time, the duration of star formation, and
the crossing time are comparable. 
This implies that partially embedded clusters may not
have enough time to form relaxed, virial clusters; this in
turn may explain in part the range of morphologies discussed in 
Section 3.



\section{EFFECTS OF CLUSTERS ON STAR AND PLANET FORMATION}

The radiation fields produced by the cluster environment can have an
important impact on stars and planets formed within. Both the extreme, 
ionizing UV (EUV) and the far-UV (FUV) 
radiation can drive disk evaporation ({\it Shu et
al.}, 1993; {\it Johnstone et al.}, 1998; {\it St{\"o}rzer and Hollenbach}, 1999;
{\it Armitage}, 2000). In the modest sized clusters of interest here
(100-1000 stars), the mass loss driven by FUV radiation generally
dominates (e.g., {\it Adams et al.}, 2004), although EUV photoevaporation can
also be important ({\it Armitage}, 2000; {\it Johnstone et al.}, 1998; {\it Shu et
al.}, 1993; {\it St{\"o}rzer and Hollenbach}, 1999).  For clusters 
with typical cluster membership e.g., with $N_{star}$ = 300 (Section 3.1),
the average solar system is exposed to a FUV flux of $G
\approx 1000 - 3000$ ({\it Adams et al.}, 2006), where $G$ = 1 corresponds to
a flux of 1.6 $\times 10^{-3}$ erg cm$^{-2}$ s$^{-1}$.  FUV fluxes of
this magnitude will evaporate a disk orbiting a solar type star down
to a truncation radius of about 50 AU over a time scale of 4 Myr. As a
result, planet forming disks are relatively immune in the regions
thought to be relevant for making giant gaseous planets. Forming solar
systems around smaller stars are more easily evaporated for two
reasons. First, the central potential well is less deep, so the
stellar gravity holds less tightly onto the disk gas, which is more
easily evaporated. Second, we expect the disk mass to scale linearly
with stellar mass so that disks around smaller stars have a smaller
supply and can be evaporated more quickly. With these disadvantages, M
stars with 0.25 $M_\odot$ can be evaporated down to 10 AU in 4 Myr
with an FUV radiation field of $G$ = 3000.
In larger clusters with more massive stars, {\it Adams et al.}  
(2004) find that regions with strong FUV and EUV can affect disks around solar mass stars 
on solar system size scales, truncating an initially 100~AU disk to a
radius of 30 AU in 4~Myr.  

A full assesment of the importance of UV radiation on disks needs to
be informed by the observed properties of clusters.  
What fraction of stars in the {\it Spitzer} and
2MASS samples are found in clusters with significant EUV radiation
fields?  We use the presence of an HII region as an indicator of a EUV
field.  In the {\it Spitzer} sample (the Orion A, Orion B and Ophiuchus
clouds) the two clusters with HII regions contain 45\% of the IR-excess 
sources.  In the 2MASS sample (Orion A, Orion B, Perseus and
Monocerous R2), 55\% of the young stars are found in the four clusters
with HII regions.  Thus, a significant fraction of stars is found in
clusters with HII regions.  However, in both the 2MASS and {\it Spitzer}
samples most of the stars found in clusters with HII regions are
found in the ONC.  The ONC has a radius of 4~pc and many of the low
mass stars in this cluster are more than a parsec away from the
massive stars, which are concentrated in the center of the cluster.
Thus, the fraction of stars exposed to a significant EUV field appears
to be less than 50\%. However, a more systematic determination of this 
fraction should be made as data become available.  

In addition to driving photoevaporation, EUV radiation (and X-rays) can
help ionize the disk gas. This effect is potentially important. One of
the most important mechanisms for producing disk viscosity is through
magneto-rotational instability (MRI), and this instability depends on
having a substantial ionization fraction in the disk. One problem with
this idea is that the disk can become too cold and the ionization
fraction can become too low to sustain the turbulence. If the
background environment of the cluster provides enough EUV radiation,
then the cluster environment can be important for helping drive disk
accretion.

Clusters can also have an affect on the processes of star and planet
formation through dynamical interactions.  This raises a variant of the  
classic question of nature vs nurture: are the properties of
the protostars and the emergent stars influenced by interactions or are they
primarily the result of initial conditions in a relatively isolated
collapse?  The numerical simulations of cloud collapse and cluster
formation ({\it Bate et al.}, 2003; {\it Bonnell et al.}, 2003;
{\it Bonnell et al.}, 2004) predict that interactions are important,
with the individual protostars competively accreting gas from a common
reservoir as they move through the cloud, and dynamical interactions  
between protostars resulting in ejections from the cloud.

We assess the importance of interactions given our current
understanding of cluster structure.  The density of clusters, and of
protostars in clusters, suggest that if stars move with velocities
similar to the turbulent gas velocity (${\rm \sim 1~kms^{-1}}$),  
interactions can occur in the lifetime of a protostar (100,000 yr).
{\it Gutermuth et al.} (2005) found typical stellar densities of 10$^4$ stars
pc$^{-3}$ in the cores of two young clusters.  If the velocity
dispersion is ${\rm 1~kms~^{-1}}$, most protostars will pass
within 1000 AU - the size of a protostellar envelope - of another star
or protostars within a protostellar lifetime.  The observed spacing of
Class 1/0 sources discussed in Section~3.2 also suggests that
interactions can occur in some cases.  At these distances protostars
could compete for gas or interact through collisions of their
envelopes.  
Interestingly, recent data
suggest that, at least in some clusters, the observed pre-stellar
clumps that make up the initial states for star formation are not
moving dynamically, but rather have subvirial velocities ({\it Walsh et
al.}, 2004; {\it Peretto et al.},  2006).  If these clusters are typical, then
interactions between protostars in clusters would be minimal.

Given the observed surface densities of clusters, is it possible that
a cluster could result from the collapse of individual,
non-interacting pre-stellar cores (i.e. nature over
nurture)?  If the starting density profile of an individual star
formation event can be modeled as an isothermal sphere, then its
radial size would be given by $r = G M_\ast / 2 a^2 \approx 0.03$ pc
(where we use a typical stellar mass of $M_\ast = 0.5 M_\odot$ and
sound speed ${\rm a = 0.2~kms~^{-1}}$). A spherical volume of radius $R$ = 1 pc
can thus hold about 37,000 of these smaller spheres (in a close-packed
configuration).  Thus, we can conclude that there is no {\it a priori}
geometrical requirement for the individual star forming units to
interact.

Once a star sheds or accretes its protostellar envelope, direct
collisions are relatively rare because their cross sections are
small. Other interactions are much more likely to occur because they
have larger cross sections. For example, the disks around newly formed
stars can interact with each other or with passing binaries and be
truncated ({\it Kobayashi and Ida}, 2001; {\it Ostriker}, 1994). In rough terms, these 
studies indicate that a    
passing star can truncate a circumstellar disk down to a radius $r_d$
that is one third of the impact parameter.
In addition, newly formed planetary systems can
interact with each other, and with passing binary star systems, and
change the planetary orbits ({\it Adams and Laughlin}, 2001). In a similar
vein, binaries and single stars can interact with each other, exchange
partners, form new binaries, and/or ionize existing binaries ({\it McMillan
and Hut}, 1996; {\it Rasio et al.}, 1995).

To affect a disk on a solar system (40 AU) scale requires a close
approach at a distance of 100 AU or less.  {\it Gutermuth et al.} (2005) 
estimated the rate of such approaches for the dense cores of
clusters.  They estimate that for the typical density of $10^4$ stars
per pc$^{-3}$, the interaction time is $10^7$ years, longer than the
lifetime of the cluster.  For N-body models of the modest sized
clusters of interest here (100-1000 members), the typical star/disk
system is expected to experience about one close encounter within 1000
AU over the next $\sim5$ Myr while the cluster remains intact; close
encounters within 100 AU are rare (e.g., {\it Adams et al.}, 2006; {\it Smith and  
Bonnell}, 2001).  Given that lifetime of the cluster is less than 5~Myr,
these models again indicate a minimal effect on nascent solar systems.

\section{\textbf{CONCLUSIONS}} 

\medskip
\noindent 
{\it 1. The Distribution of Cluster Properties:} Systematic surveys of
giant molecular clouds from 2MASS and {\it Spitzer}, as well as targeted
surveys of individual clusters, are providing the first measurements
of the range and distribution of cluster properties in the nearest
kiloparsec.  Although most stars appear in groups or clusters, in many star--forming 
regions there is a significant distributed component.  
These results suggest that there is 
a continuum of star--forming environments from
relative isolation to dense clusters.  Theories of star formation must
take into account (and eventually explain) this observed distribution.
The 2MASS and {\it Spitzer} surveys also show a 
correlation between number of member stars and the radii of
clusters, such that the average surface density of stars varies  
by a factor of only $\sim$5. 

\medskip
\noindent 
{\it 2. The Structure of Young Stellar Clusters:} 
Common cluster morphologies include
elongation, low density halos, and sub-clustering. 
The observed cluster and molecular gas morphologies are similar, 
especially when only the youngest Class I/0 sources are considered. 
This similarity suggests that these morphologies (except possibly halos) 
result from the distribution of
fragmentation sites in the parental cloud, and not the subsequent
dynamical evolution of the cluster. 
Consequently, the surface densities and
morphologies of clusters are important constraints on models of the
birth of clusters.

\medskip
\noindent 
{\it 3. The Evolution of Clusters:} The evolution of clusters is driven
initially by the formation of stars, and then later on by the
dissipation of gas.  Gas dissipation appears to be driven by different
processes in different regions, including photoevaporation by  
extreme-UV from O stars, photoevaporation by far-UV radiation from B
stars, and outflows from lower mass stars.  Much of the gas appears to
be dissipated in 3~Myr which is a few times the crossing time and the  
duration of star formation in these clusters.  With these short timescales, 
clusters probably never
reach dynamical equilibrium in the embedded phase.  The survival of
clusters as the gas is dispersed is primarily a function of the size
of the cluster, the efficiency of star formation, and the rate at
which the gas is dispersed.

\medskip
\noindent 
{\it 4. The Impact of Clustering on Star and Planet Formation:} Far-UV and
Extreme-UV radiation from massive stars can effectively truncate disks
in a few million years.  Extreme UV radiation is needed to affect
disks around solar type stars on solar system scales ($< 40$~AU) in the lifetime of the
cluster.  Within our sample of molecular clouds, fewer than 50\%  
of the stars are found in regions with strong extreme UV-fields.  The
observed spacing of protostars suggest that dynamical interactions and
competitive accretion may occur in the denser regions of the observed
clusters.  However, evidence of sub-virial velocities of pre-stellar
condensations in at least one cluster hints that these
interactions may not be important.
Given the densities and lifetimes of the observed
clusters, dynamical interactions do not appear to be an important
mechanism for truncating disks on solar system size scales. 

\bigskip 
\textbf{ACKNOWLEDGMENTS} 
This work is based in part on observations made with the {\it Spitzer Space Telescope}, which is
operated by the Jet Propulsion Laboratory, California Institute of Technology under
NASA contract 1407. Support for this work was provided by NASA through Contract Numbers  
1256790 and 960785, issued by JPL/Caltech. 
PCM acknowledges a grant from the {\it Spitzer} Legacy Science Program to the ``Cores to Disks" team 
and a grant from the NASA Origins of Solar Systems Program.
S.J.W. received support from {\it Chandra} X-ray Center contract NAS8-39073. 
FCA is supported by NASA through the Terrestrial Planet Finder Mission
(NNG04G190G) and the Astrophysics Theory Program (NNG04GK56G0). 

\bigskip
\centerline\textbf{ REFERENCES}
\bigskip
\parskip=0pt
{\small
\baselineskip=11pt

\refs Adams F. C. (2000)  {\it Astrophys. J., 542}, 964-973.

\refs Adams F. C. and Laughlin G. (2001) {\it Icarus, 150}, 151-162.

\refs Adams F. C., Lada C. J., and Shu F. H. (1987) {\it Astrophys. J., 312}, 788-806.

\refs Adams F. C., Hollenbach D., Laughlin G., and Gorti U. (2004) {\it   Astrophys. J., 611}, 360-379.

\refs Adams F. C., Proszkow E. M., Fatuzzo M., and Myers P. C.
(2006) {\it Astrophy. J., in press}, astro-ph/0512330

\refs Allen L. E., Myers P. C., Di Francesco J., Mathieu R.,
Chen H., and Young E. (2002) {\it Astrophys. J., 566}, 993-1004.

\refs Allen L. E., Calvet N., D'Alessio P., Merin B., Hartmann L., Megeath S. T., et al. (2004) {\it Astrophy. J. Suppl., 154}, 363-366. 

\refs Allen L. E., Hora J. L., Megeath S. T., Deutsch L. K., Fazio G. G., Chavarria L., and Dell R. D. 
(2005) {\em Massive Star Birth: A Crossroads of Astrophysics} (R. Cesaroni, E. Churchwell, M. Felli, and C. M. Walmsley eds.), 
pp. 352-357. Cambridge University Press, Cambridge.

\refs Allen L. E., Harvey P., Jorgensen J., Huard T., Evans N. J. II et al.  (2006) {\it in prep.}

\refs Andr\'e P., Ward-Thompson D., and Barsony  M. (1993) {\it Astrophys. J., 406}, 122-141.

\refs Andr\'e P., Ward-Thompson D., and Barsony  M. (2000) {\it Protostars and Protoplanets IV} (V. Mannings, A. Boss, S. Russell, eds.), pp. 59-96. University of Arizona, Tucson.
 
 
\refs Armitage P. J. (2000) {\it Astron. Astrophys., 362}, 968-972.
 
 
\refs Bally, J., Stark A. A., Wilson R. W., and Langer W. D. (1987) {\it Astrophys. J., 312}, L45-L49. 

\refs Bate M. R., Bonnell I. A., and Bromm V. (2003) {\it Mon. Not. R. Astron. Soc., 339}, 577-599.
 
\refs Boily C. M. and Kroupa P. (2003a) {\it Mon. Not. R. Astron. Soc., 338}, 665-672.
 
\refs Boily C. M. and Kroupa P. (2003b) {\it Mon. Not. R. Astron. Soc., 338}, 673-686.
 
 
 
\refs Bonnell I. A., Bate M., and Vine S. G. (2003) {\it Mon. Not. R. Astron. Soc., 343}, 413-418.
 
\refs Bonnell I. A., Vine S. G., and Bate M. (2004) {\it Mon. Not. R. Astron. Soc., 349}, 735-741.

\refs Cambr\'esy L., Petropoulou V., Kontizas M., and Kontizas E. (2006) {\it Astron. Astrophys., 445}, 999-1003.

\refs Carpenter J. M., Meyer M. R.,
Dougados C., Strom S. E., and Hillenbrand L. A. (1997) {\it Astron. J., 114}, 198-221.

\refs Carpenter J. M. (2000) {\it Astron. J., 120}, 3139-3161.

\refs Carpenter J. M., Hillenbrand L. A., and Skrutskie M. F. (2001) {\it Astron. J., 121}, 3160-3190.

\refs Carpenter J. M., Hillenbrand L. A., Skrutskie M. F., and Meyer M. R. (2002) {\it Astron. J., 124}, 1001-1025.

\refs Casassus S., Bronfman L., May J., and Nyman L. A. (2000) {\it Astron. Astrophys., 358}, 514-520.

\refs Chen H., Tafalla M., Greene T. P., Myers P. C., and Wilner D. J. (1997) {\it Astrophys. J., 475}, 163-172.


\refs Comeron F. and Torra J. (1996) {\it Astron. Astrophys., 314}, 776-784.


\refs Dorren J. D., Guedel M., and Guinan E. F. (1995) {\it Astrophys. J., 448}, 431-436.  

\refs Evans N. J. II, Allen L. E., Blake G. A., Boogert A. C. A., Bourke T. et al. (2003) {\it Publ. Astron. Soc. Pac., 115}, 965-980.


\refs Feigelson E. D. and Montmerle T. (1999) {\it Ann. Rev. Astron. Astrophys., 37}, 363-408.

\refs Feigelson E. D, Getman K., Townsley L., Garmire G., Preibisch T., Grosso N., Montmerle T., Muench A., and McCaughrean M. (2005) {\it Astrophys. J. Suppl., 160}, 379-389.

\refs Flaccomio E., Damiani F., Micela G., Sciortino S., Harnden F. R., Murray S. S., and Wolk S. J. (2003) {\it Astrophys. J., 582}, 398-409.

\refs Flaherty K., Pipher J. L., Megeath S. T., Winston E. A., Gutermuth R. A., and 
Muzerolle J.  (2006) {\it in prep.}



\refs Gladwin P. P., Kitsionas S., Boffin H. M. J., and Whitworth A. P. (1999) {\it Mon. Not. R. Astron. Soc., 302}, 305-313. 


\refs Gomez M., Hartmann L., Kenyon S. J., and Hewett R. (1993) {\it Astron. J., 105}, 1927-1937.

\refs Greene T. P., Wilking B. A., Andr\'e, P., Young, E. T., and Lada C. J. (1994) {\it Astrophys. J., 434}, 614-626. 




\refs Gutermuth R. A. (2005) {\it Ph.D. thesis}, Univ. of Rochester.

\refs Gutermuth R. A., Megeath S. T., Muzerolle J., Allen L. E., Pipher J. L., Myers P. C., and Fazio G. G. (2004) {\it Astrophys. J. Suppl., 154}, 374-378.

\refs Gutermuth R. A., Megeath S. T., Pipher J. L., Williams J. P., Allen L. E., Myers P. C., and Raines S. N. (2005) {\it Astrophys. J., 632}, 397-420.

\refs Gutermuth R. A., Pipher J. L., Megeath S. T., Allen L. E., Myers P. C.,  et al. (2006) {\it in prep.}

\refs Haisch K. E., Lada E. A., and Lada C. J. (2001) {\it Astrophys. J. 553}, L153-L156.

\refs Hartmann L. (2004) In {\it Star Formation at High Angular Resolution} (M. Burton, R. Jayawardhana, T. Bourke, eds.), pp. 201-211. ASP Conf. Series, San Francisco. 

\refs Hartmann L., Megeath S. T., Allen L., Luhman K., Calvet N. et al. (2005) {\it  Astrophys. J., 629},  881-896.

\refs Herbig G. H. and Dahm S. E. (2002) {\it Astron. J. 123}, 304-327.

\refs Herbig G. H. and Bell K. R. (1998) {\it Lick Observatory Bulletin, Santa Cruz: Lick Observatory,
1995}, VizieR Online Data Catalog, 5073. 

\refs Hillenbrand L. A. (1995)  {\it Ph.D. Thesis, University of Massachusetts}.

\refs Hillenbrand L. A., Strom S. E., Vrba F. J., and Keene J. (1992) {\it  Astrophys. J., 397}, 613-643.

\refs Hillenbrand L. A., Massey P., Strom S. E., and Merrill K. M. (1993) {\it Astron. J., 106}, 1906-1946.

\refs Hillenbrand L. A. and Hartmann L. (1998) {\it Astrophys. J., 492}, 540.

\refs Horner D. J., Lada E. A., and Lada C. J. (1997) {\it Astron. J. 113}, 1788-1798.

\refs Huard T. (2006) {\it private communication.}

\refs Johnstone D., Hollenbach D. J., and Bally J. (1998) {\it Astrophys. J., 499}, 758-776.

\refs Kaas A. A. (1999) {\it Astron. J., 118}, 558-571.

\refs Kaas A. A., Olofsson G., Bontemps S., Andr\'e P., Nordh L. et al. (2004) {\it Astron. Astrophys. 421}, 623-642.




\refs Kobayashi H. and Ida S. (2001) {\it Icarus, 153}, 416-429.




\refs Lada E. A., Evans N. J. II, Depoy D. L., and Gatley I. (1991) {\it Astrophys. J., 371}, 171-182.

\refs Lada C. J. and Lada E. A. (1995) {\it Astron. J., 109}, 1682-1696.

\refs Lada C. J., Alves J., and Lada E. A. (1996) {\it Astron. J., 1111}, 1964-1976.

\refs Lada C. J., Muench A. A., Haisch K. E., Lada E. A., Alves J. F., Tollestrup E. V., and  Willner S. P. (2000) {\it Astron. J., 120}, 3162-3176.

\refs Lada C. J. and Lada E. A. (2003) {\it Ann. Rev. Astron. Astrophys., 41}, 57-115.

\refs Lada C. J., Margulis M., and Dearborn D. (1984), {\it Astrophys. J., 285}, 141-152.


\refs Larson R. B. (1985) {\it Mon. Not. R. Astron. Soc., 214}, 379-398.

\refs Leisawitz D., Bash F. N., and Thaddeus P. (1989) {\it Astrophys. J. Suppl., 70}, 731-812.

\refs Luhman K. L., Stauffer J. R., Muench A. A., Rieke G. H., Lada E. A., Bouvier J., and Lada C. J.
(2003) {\it Astrophys. J., 593}, 1093-1115.

\refs Matzner C. D. and McKee C. F. (2000) {\it Astrophys. J., 545}, 364-378.

\refs McMillan S. L. W. and Hut P. (1996) {\it Astrophys. J., 467}, 348-358.

\refs Megeath S. T., Herter T., Beichman C., Gautier N., Hester J. J., Rayner J., and Shupe D. (1996)
{\it Astron. Astrophys., 307}, 775-790.

\refs Megeath S. T. and Wilson T. L. (1997) {\it Astron. J., 114}, 1106-1120.

\refs Megeath S. T., Biller B., Dame, T. M., Leass E., Whitaker R. S., and Wilson T. L.  (2002) {\it Hot Star Workshop III: The Earliest Stages of MAssive Star Formation} (P. A. Crowther, ed.), pp. 257-265. ASP Conf. Series, San Francisco. 

\refs Megeath S. T., Allen L. E., Gutermuth R. A., Pipher J. L., Myers P. C. et al. (2004) {\it
Astrophys. J. Suppl., 154}, 367-373.

\refs Megeath S. T., Flaherty K. M., Hora J., Allen L. E., Fazio G. G. et al. (2005) {\it Massive Star Birth: A Crossroads of Astrophysics} (R. Cesaroni, E. Churchwell, M. Felli, and C. M. Walmsley eds.), pp. 383-388. Cambridge University Press, Cambridge.

\refs Megeath S. T., Flaherty K., Gutermuth R., Hora J., Allen L. E. et al. (2006) {\it in prep.}


\refs Micela G.,  Sciortino S., Serio S., Vaiana G.~S., Bookbinder J., Golub L., Harnden F.~R.,
and Rosner R. (1985) {\it Astrophys. J., 292}, 172-180.

\refs Miesch M. S. and Bally J. (1994) {\it Astrophys. J., 429}, 645-671. 

\refs Morris P. W., Noriega-Crespo A., Marleau F. R., Teplitz H. I., Uchida K. I., and Armus L. (2004) {\it Astrophys. J. Suppl., 154}, 339-343.

\refs Motte F., Andr\'e P. and Neri R. (1998) {\it Astron. Astrophys., 336}, 150-172. 

\refs Motte F., Andr\'e P., Ward-Thompson D., and Bontemps S. (2001) {\it Astron. Astrophys., 372}, L41-L44.

\refs Muench A. A., Lada E. A., Lada C. J., Elston R. J., Alves J. F., Horrobin M., Huard T. H. et al.
(2003) {\it Astron. J., 125}, 2029-2049.

\refs Muzerolle J., Megeath S. T., Gutermuth R. A., Allen L. E., Pipher J. L. et al. (2004) {\it Astrophys. J. Suppl., 154}, 379-384.

\refs Muzerolle J., Megeath S. T., Flaherty K., Allen L. E., Young E. T. et al. (2006) {\it in prep.} 

\refs Myers P. C. and Ladd E. F. (1993) {\it Astrophys. J., 413}, L47-L50.


\refs Nutter D. J., Ward-Thompson D., and Andr\'e P. (2005) {\it Mon. Not. R. Astron. Soc., 357},  975-982. 

\refs Ostriker E. C. (1994) {\it Astrophys. J., 424}, 292-318.

\refs Peretto N., Andr\'e P., and Belloche A. (2006) {\it Astron. Astrophys., 445}, 879-998. 

\refs Porras A., Christopher M., Allen L., Di Francesco J., Megeath S. T., and Myers P. C. (2003) {\it Astron. J., 126}, 1916-1924.


\refs Quillen A. C., Thorndike S. L., Cunninghman A., Frank A., Gutermuth R. A., Blackmann E. G.,
Pipher J. L, and Ridge N. (2005) {\it Astrophys. J., 632}, 941-955.

\refs Rasio F. A., McMillan S., and Hut P. (1995) {\it Astrophys. J., 438}, L33-L36.


\refs Rodr\'iguez L. F., Anglada G., and Curiel S. (1999) {\it Astrophys. J. Suppl., 125}, 427-438.

\refs Sandell G. and Knee L. B. G. (2001) {\it Astrophys. J., 546},  L49-L52.

\refs Shu F. H., Johnstone D., and Hollenbach D. J. (1993) {\it Icarus,
106}, 92-101.

\refs Smith K. W. and Bonnell I. A. (2001) {\it Mon. Not. R. Astron. Soc., 322}, L1-L4.

\refs Stern D., Eisenhardt P., Gorjian V., Kochanek C. S., and Caldwell N. (2005) {\it Astrophys. J., 631}, 163-168. 


\refs St\"orzer H. and Hollenbach D. (1999) {\it Astrophys. J., 515}, 669-684.


\refs Teixeira P. S., Lada C. J., Young E. T., Marengo M.,
and Muench A., et al. (2006) {\it Astrophys. J., 636}, L45-L48. 

\refs Testi L. (2002) In {\it Modes of Star Formation and the Origin
of Field Populations} (E. K. Grebel
and W. Brandner, eds.), pp. 60-70. 

\refs Walsh A. J., Myers P. C., and Burton M. G. (2004) {\it Astrophys. J., 614}, 194-202.

\refs Walter F. M. and Barry D. C. (1991) {\it The Sun in Time}, 633-657.

\refs Whitney B. A., Wood K., Bjorkman J. E., and Cohen M. (2003) {\it Astrophys. J., 598}, 1079-1099.


\refs Wilking B. A., Schwartz R. D., and Blackwell J. H. (1987) {\it Astron. J.,  94}, 106-110.

\refs Wilking B. A., Lada C. J., and Young E. T. (1989) {\it Astrophys. J., 340}, 823-852.

\refs Wiramihardja S. D., Kogure T., Yoshida S., Ogura K., and
Nakano M. (1989) {\it Planet. Astron. Soc. Pac., 41}, 155-174.

\refs Wolk S. J., Spitzbart B.D., Bourke T.L., and Alves J. (2006) {\it Astron. J., submitted}. 

\refs Young E. T., Teixeira P., Lada C. J., Muzerolle J., Persson S. E. et al. (2006) {\it Astrophys. J., in press}, astro-ph/0601300. 
}

\end{document}